\documentclass[aps,prl,twocolumn,showpacs]{revtex4-1}
\usepackage{graphicx}
\usepackage{dcolumn}
\usepackage{bm}
\usepackage{amssymb}
\begin{document}
\title{Order-Parameter Anisotropies in the Pnictides -
An Optimization Principle for Multi-Band Superconductivity}

\author{Christian Platt${}^1$}
\author{Ronny Thomale${}^2$}
\author{Werner Hanke${}^1$}
\affiliation{${}^1$Institute for Theoretical
  Physics and Astrophysics, University of W\"urzburg, Am Hubland, D
  97074 W\"urzburg}
\affiliation{${}^2$Department of Physics, Princeton University, Princeton,
  NJ 08544}

\begin{abstract}
Using general arguments of an optimization taking place between the
pair wave function and the repulsive part of the electron-electron
interaction, we analyze the superconducting gap in materials with
multiple Fermi-surface (FS) pockets, with exemplary application to
two proto-type ferropnictide setups. On the basis of functional
renormalization group (FRG) calculations for a wide parameter span
of the bare interactions, 
%and for the different FS topologies
%applying to these two characteristic Fe-based superconductors, 
we show that the symmetry of the gap and the nodal versus nodeless
behavior is driven by this optimization requirement.
\end{abstract}
\date{\today}

\maketitle

The discovery of superconductivity (SC) in new compounds such as the
iron pnictides has opened up a new avenue for studying the fundamental
question "what is universal and what is material-dependent" concerning
the mechanism of high-$T_{\text{c}}$ SC in a wider class of materials
other than, but also including, the cuprates. There, after more than
two decades of intense research, more and more theoretical as well as
experimental studies support a scenario where the general nature of
the $d$-wave SC as well as other salient features are accounted for by
an electronic pairing mechanism extracted from a one-orbital Hubbard
model~\cite{Anderson:1987, Scalapino:2007} with the addition that the
material-dependence is embedded in the multi-orbital (e.g. 3-band)
extensions \cite{Hackl:2010}.

In other SC compounds such as the pnictides, however, the picture
seems more complicated: Here, at the outset, multi-band SC appears
with gaps possibly displaying different symmetries such as extended
(sign-reversing) $s$- competing with $d$-wave and with nodal or also
nodeless behavior on the disconnected Fermi surface (FS) sheets
\cite{Mazin:2008,Kuroki:2008,Graser:2009,Stanev:2008,Chubokov:2008,Wang:2009,Platt:2009,Thomale:2009,Zhai:2009,Wang:2010,Chubokov:2009,Thomale:2010,Kemper:2010,
  Maiti:2010,Maier:2010}. Accordingly, even the simplest multi-band
Hamiltonian with only on-site interactions contains four
possibly relevant terms, the intra-orbital and inter-orbital
repulsion as well as the Hund's-rule coupling and pair hopping.
Searching for SC pairing, these interactions have to be augmented with
the orbital dependence of the FS pockets, since the interactions
become matrices formed by local orbitals which have a dominant
orbital "weight" at the FS pockets. We hence investigate whether this
intricate interplay of multi-orbital band structure, FS topology and
interactions still allows for insights into a more universal than
material-dependent understanding of SC in these systems.

In this Letter, we describe an "optimization principle" which can help
providing such a more universal picture. The main point of our work is
to show that the SC state, its gap, and, in particular, its anisotropy
in momentum space is determined by an optimization \textit{which
  determines and optimizes the interplay between the attractive interaction in the
  SC-channel and the Coulomb repulsion.} This optimization problem, as
discussed below, is unavoidable in a multi-band SC situation: for the
pnictides, it appears because of a frustration in the
$s_{\pm}$-channel, when more than two FS-pockets are involved in
setting up the pairing interaction.

%A similar set of two FS
%topologies and its relation to gap anisotropies has also been
%studied by Maiti and Chubukov in a recent analytical RG work
%\cite{Maiti:2010} and by Maier et al.~\cite{Kemper:2010, Maier:2010}
%and Kuroki et al.~\cite{Kuroki:2008} in spin-fluctuation models.
%Both the 4pFS and the 5pFS are believed to have material
%representatives in the pnictide family, with the latter related to
%$LaFeAsO$ and the former to $LaFePO$, where the difference in band
%structure essentially stems from the varying pnictogen height
%replacing As by P~\cite{Kuroki:2008,Thomale:2010}. In particular,
%building on our previous work, we demonstrate that this principle
%can explain why a fully-gapped sign-reversing $s$-wave SC is favored
%in the generic 5pFS, while a nodal pairing is expected for 4pFS,
%consistent with the majority of experiments (see, for example Ref.
%\cite{Kim:2010}). The calculations displaying the optimization
%principle for these two proto-type setups, i.e. 4pFS and 5pFS, are
%performed using the functional renormalization group (FRG).

Already from the BCS gap equation, one can see that Coulomb
repulsion at a finite momentum transfer can induce pairing only when
the wave vector of such an interaction connects regions on one FS
(in the cuprate case), or regions on different FSs (in the pnictide
case), which have opposite signs of the SC order parameter. This
corresponds to putting the electron pairs in an anisotropic wave
function such as $d$-wave in the high-$T_{\text{c}}$ cuprates, or the
sign-reversing  $s$-wave ($s_{\pm}$) in the pnictides, where in the
latter case the wave vector ($\pi,0$) in the unfolded Brillouin Zone
connects hole (h) and electron (e) FS-pockets with a sign-changing
$s_{\pm}$ gap~\cite{Mazin:2008,Chubokov:2008}. Early studies based
on either RPA spin-fluctuations (SF) scenarios~\cite{Mazin:2008} or
on Renormalization-Group [RG] studies~\cite{Chubokov:2008} of just
one-hole and one-electron FS have reported a momentum-independent
$s_{\pm}$ gap. At first glance, this similarity of the gap function
obtained by so dissimilar approaches as RPA and FRG may appear
surprising. Indeed, the repulsive part of the Coulomb interaction is
treated differently which leads to differing results for the general
multi-pocket
case~\cite{17}.

%The general aspect of the orthogonality scenario is similar to a
%discussion presented by Anderson in a recent article
%\cite{Anderson:2007}, where he argues that pairing in conventional
%"low-$T_{\text{c}}$" SC has a rather different microscopic origin from the
%high-$T_{\text{c}}$ cuprates and many other unconventional SC. His point is,
%that the strongly repulsive (short-range) part of the Coulomb
%interaction is avoided by choosing the SC pair state orthogonal to
%the repulsive core of the Coulomb interaction, i.e. putting the
%electron pairs in an anisotropic wave function (such as $d$-wave),
%which vanishes at the core of the Coulomb interaction (see also Ref.
%\cite{Scalapino:1995}, Eq. (19)). The "$s_{\pm}$" superconducting
%state, early on conjectured by Mazin et al.~\cite{Mazin:2008}, which
%is consistent with experimental observations for the Fe-based SC,
%may be viewed as a specific version of this argument.

\begin{figure*}[t]
  \begin{minipage}[l]{0.85\linewidth}
%    \psfrag{a}{$0$}
%    \psfrag{b}{$\uparrow$}
%    \psfrag{c}{$\downarrow$}
%    \psfrag{D}{\small$\Delta_3$}
    \includegraphics[width=\linewidth]{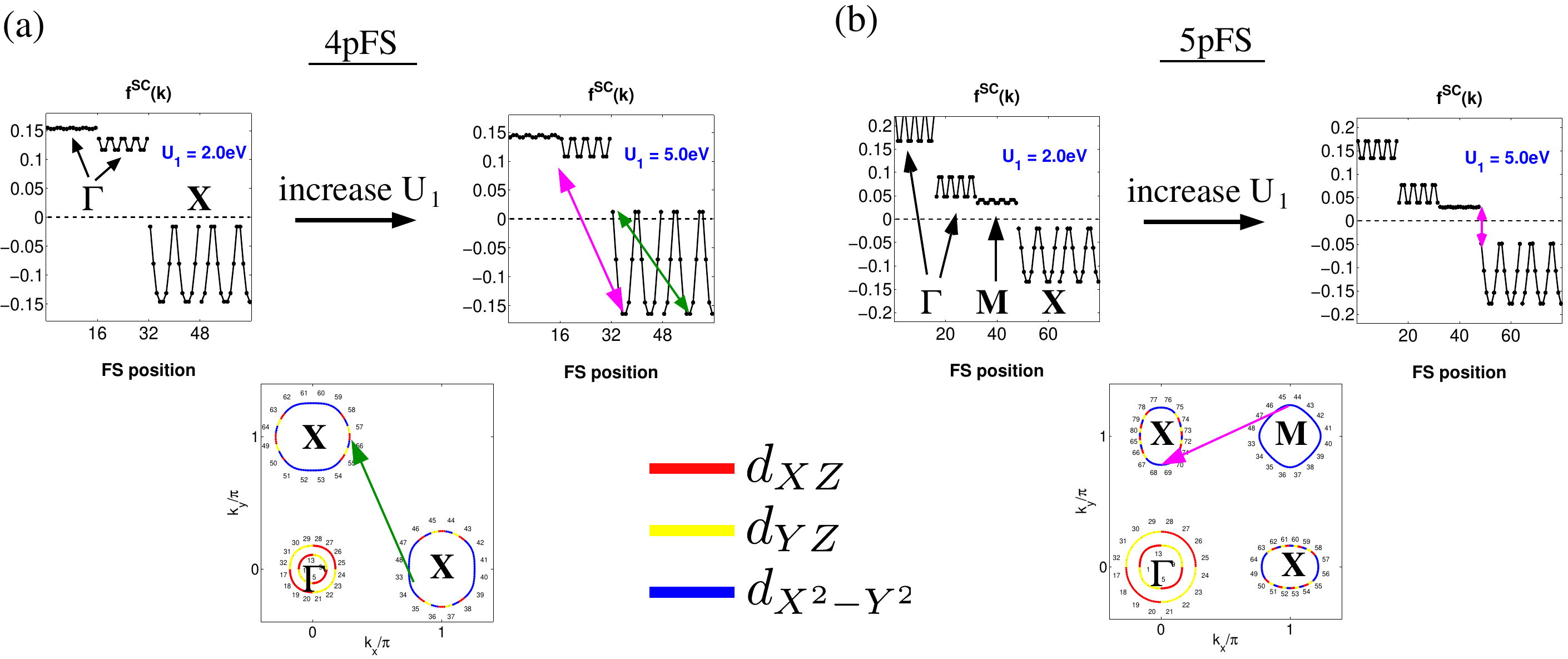}
  \end{minipage}
  \caption{(Color online). Scan of the intra-orbital interaction $U_{1}$ for both the 4pFS (Fig. \ref{su-pic1}a) and 5pFS (Fig. \ref{su-pic1}b) situations. Colors on the FS-pockets give dominant orbital weights, with dashed segments signalling several
leading contributions. The increase of $U_{1}$ leads to an increase of
the e-pocket gap for the 5pFS scenario, while it favors the nodal
scenario for 4pFS.}
\label{su-pic1}
\vspace{-0pt}
\end{figure*}

The interesting setup for the optimization principle concerns a
multi-pocket situation - as generally appearing in the ferro pnictides
- where more than two pockets create crucial pairing interactions. In
order to illustrate the principle at work for such a scenario, we
investigate a 4-pocket and a 5-pocket Fermi surface (pFS) topology
originating from a 5-band model (Fig. \ref{su-pic1}), and discuss the
superconducting order parameter from that perspective.
%As shown with a blue arrow in Fig. \ref{su-pic1}, this occurs e.g.
%in the P-based compounds between the two e-pockets centered at the
%$X(\pi, 0)$-point, and in the As-based compounds, as indicated by a
%red arrow, between the e-pockets (X) and an additional h-pocket (M)
%at $(\pi, \pi)$ in the unfolded Brillouin Zone. It is obvious that
%this multi-pocket situation induces a rather complex competition and
%, in particular, also frustration with the above $s_{\pm}$ scenario:
%for example, in Fig. \ref{su-pic1}a), in the P-based situation, the
%again repulsive (blue) interaction between the X-pockets tends to
%also push the two e-pockets into a sign-reversing $s_{\pm}$
%situation, eventually driving the gap on these e-pockets from a
%nodeless to a nodal behavior. However, this latter nodal behavior
%obviously creates a frustration with the, also present,
%sign-reversed gap tendency between h- and e-pockets. How can the
%system then strike a compromise? This is indicated in the RG
%calculations of Fig. \ref{su-pic1} with the additional red arrow in
%the gap function $f^{SC}(k)$ versus momentum $k$ plots (in Figs.
%\ref{su-pic1}a). It gives the dominant (dependent on the orbital
%weights on the FS pockets) interaction between h- and e-pockets. It
%obviously acts so as to increase the anisotropy by pushing the
%"valleys" of the e-pocket gaps further down to increase the
%corresponding $s_{\pm}$ tendency.

The principal physical content of this optimization scenario can
already be observed for the 4pFS and 5pFS scenarios in
Figs. \ref{su-pic1}a and b. Let us try to understand the systems from
the unfrustrated $s_\pm$ limit where the $\Gamma \leftrightarrow X$
pair scattering between h-pockets at $\Gamma (0,0)$ and 3-pockets
at $X (\pi,0)$ is minimized. Here, a green arrow for $X \leftrightarrow X$
scattering (4pFS, lower FS display in Fig.~\ref{su-pic1}a) and a magenta arrow (5pFS) for $X \leftrightarrow M$ 
indicate additional interactions (dependent on the dominant orbital weights on the FSs),
%which appear between two e-pockets at the $X(\pi, 0)$-points (4pFS)
%and, additionally between the $X$-pockets and a h-pocket at the
%$M(\pi, \pi)$-point (5pFS). 
The green interaction in
Fig. \ref{su-pic1}a 
%tends to push the two $X$-pockets into an
%$s_{\pm}$-situation, 
frustrates the previous pure $s_{\pm}$ limit. The system then strikes a
compromise - and this is the essence of the optimization principle -
by enhancing the anisotropy of the gap function (denoted by
$f^{\text{SC}}({\bf k})$ in Fig. \ref{su-pic1}) on the
e-pockets at X (FS positions 32 to 64 in Fig. \ref{su-pic1}),
eventually reaching even a nodal situation for larger
interactions. The dominant part of the $X\leftrightarrow X$ interaction  acts
so as to push the peaks of the e-gap function further up, while the dominant part of the $(\Gamma\leftrightarrow X)$
interaction (magenta arrow in Fig. \ref{su-pic1}a) tries to push the
e-gap  valleys down. 
Thus, a transparent understanding of the
anisotropies and the nodeless versus nodal behavior emerges: the
multi-band SC adjusts the momentum dependence of the gap, i.e.  its
anisotropy, so as to minimize the effect of the Coulomb repulsion~\cite{18}.
%which is due to the $s_{\pm}$-frustration
%The corresponding (again repulsive) interaction was considered by
%Mazin et al.  as spin-fluctuation (SF) induced and extracted from an
%RPA calculation. It connects well-separated FS pockets located
%around $\Gamma$-and around M-points in the Brillouin zone. Though
%repulsive in the singlet channel, the corresponding contributions
%are strongly pairing provided the order parameters (gap functions)
%on the two sets of the FS have opposite sign, i.e. form a $s_{\pm}$
%SC state. As mentioned above, this argument has to be supplemented
%with the dominant orbital dependence of the FS pockets.
%Extensive RPA-calculations considering the SF-mediated interactions
%as well as their dependence on the dominant orbital "weights" of the
%FS-pockets have been performed by Kuroki et al.~\cite{Kuroki:2008},
%by Graser et al.~\cite{Graser:2009} and in recent calculations by
%the same group~\cite{Kemper:2010}. RG calculations, taking the
%orbital weights into account, have been carried out by D.-H. Lee's
%\cite{Wang:2009,Zhai:2009,Wang:2010}, Chubukov's
%\cite{Chubokov:2008,Chubokov:2009}, and our group
%\cite{Platt:2009,Thomale:2009,Thomale:2010}.

In more mathematical terms, this optimization is reflected in Eq.~\ref{decomp}  below for the dominant Cooper-channel eigenvalue $c^{\text{SC}}_1
(\Lambda)$ taking the largest negative value:
\begin{equation}
\label{sc-ev}
c^{\text{SC}}_1(\Lambda)=\langle f^{\text{SC}}({\bf k})V^{\text{SC}}_{\Lambda}({\bf k},-{\bf k},{\bf p})f^{\text{SC}}({\bf p})^*\rangle
\end{equation}
Here, as detailed in Eq.~\ref{decomp}, $V^{\text{SC}}_{\Lambda}$ denotes the
pairing function, where $\Lambda$ is the RG-flow parameter and
$f^{\text{SC}}({\bf k})$ the SC (gap) form factor associated with it. $\langle...\rangle$ denotes the
inner product and involves the ${\bf k}$- and ${\bf p}$-points on
all 4 (or 5) FS-pockets (Fig.~\ref{su-pic1}). We have 
\begin{equation}
\label{sc-po}
c^{\text{SC}}_1(\Lambda)=\sum_{\text{FS}\
l,m}c^{\text{SC}}_{l,m}(\Lambda),
\end{equation}
and its largest negative value is
determined via an optimization taking place between all
pockets $l$ and $m$. This is a frustration problem as not all
minimization conditions can be fulfilled at the same time.

%In this work, we show that the above \textit{"orthogonality"
%argument is not only at work explaining a possible sign-change}
%between gaps on different FS sheets (such as the fully-gapped
%$s_{\pm}$-state in $LaOFeAs$) \textit{but it explains also the SC
%gap anisotropies}. These anisotropies can become very pronounced
%because of the orthogonality requirement, eventually driving the gap
%(for example, on the electron (e)-pockets in $LaOFeP$) nodal, i.e.
%having different signs on the  two e-pockets (denoted by X in Fig.
%\ref{su-pic1}(a)).

In conjunction with the underlying FS topology obtained from
LDA-type calculations~\cite{Kuroki:2008}, our FRG-studies also
serve to answer which of the four intrasite
interactions in the starting Hamiltonian are playing a leading role.
To this extent, we report on an extensive parameter-sweep study the
results of which are fully in line with the optimization argument.
Placing the value of these interactions finally around the spread of
values obtained in recent ab-initio DFT work~\cite{Imada:2010}, we
find the intra-orbital interaction $U_1$ to take on the pivotal
role. 
%Similar findings on the leading role of $U_1$ were reported in
%RPA calculations~\cite{Maier:2010}.
%Let us start with a few remarks about the FRG and why we consider
%this method especially suited for demonstrating the orthogonality
%principle and identifying the leading interactions in the
%Hamiltonian. The FRG works best for weak to intermediate
%correlations~\cite{}, an assumption we make in accordance with a
%majority of experiments on the Fe-based SC~\cite{}. This FRG
%technique allows for a controlled renormalization from the
%"high-energy" scale of the bare interactions in the starting
%Hamiltonian down to the low-energy critical diverging scale of $k_B
%T_c \sim \Lambda_c$, taking the competing fluctuations (magnetic,
%SC, screening, vertex corrections) into account.
%A two-dimensional tight-binding model spanned by the 5 Fe
%$d$-orbitals is used to describe the band structure of the 1111-type
%iron-based superconductors~\cite{Kuroki:2008}, i.e.
The 4pFS and 5pFS scenarios can be cast into a 5-band model, with
\begin{equation}
\label{h0}
H_{0} = \sum_{{\bf k},s}\sum_{a,b=1}^{5}c_{{\bf k}as}^{\dagger}K_{ab}({\bf k})c_{{\bf k}bs}^{\phantom{\dagger}}.
\end{equation}
Here $c$'s stand for electron annihilation operators, $a,b$ for the
$d$-orbitals, and $s$ denote the spin indices.  %While the main
%electronic structure of P-based and As-based compounds is very
%similar, in the vicinity of the Fermi surface, the most notable
%difference is the presence (for small h-doping) or absence of a
%broad $d_{X^2-Y^2}$ ($d_{xy}$)-dominated band at $M=(\pi,\pi)$, in
%agreement with ARPES data. To account for this difference, we use a
%{\it $5$-pocket scenario} for the {\it As-based} and a {\it
%$4$-pocket scenario} for the {\it P-based} compounds (see Fig.
%\ref{su-pic1}).

In the many-body part the intra- and inter-orbital interactions
$U_1$ and $U_2$, as well as the Hund's coupling $J_{\text{H}}$ and the pair
hopping $J_{\text{pair}}$ enter, i.e.
\begin{eqnarray}
\label{hint}
&&H_{\text{int}}=\sum_i \left[ U_1 \sum_{a} n_{i,a\uparrow}n_{i,a\downarrow} + U_2\sum_{a<b,s,s'} n_{i,as}n_{i,bs'} \right.\nonumber \\
&& \hspace{-15pt}\left.+\sum_{a<b}(J_{\text{H}} \sum_{s,s'} c_{ias}^{\dagger}c_{ibs'}^{\dagger}c_{ias'}^{\phantom{\dagger}}c_{ibs}^{\phantom{\dagger}}  +J_{\text{pair}} c_{ia\uparrow}^{\dagger}c_{ia\downarrow}^{\dagger}c_{ib\downarrow}^{\phantom{\dagger}}c_{ib\uparrow}^{\phantom{\dagger}}) \right]\hspace{-4pt},
\end{eqnarray}
where $n_{i,as}$ denote density operators at site $i$ of spin $s$ in
orbital $a$.  Typical interaction settings are dominated by
intra-orbital coupling, $U_1 > U_2 > J_{\text{H}} \sim
J_{\text{pair}}$,
% taking the values around $U_1 \cong 3.5 eV, U_2
%\cong 2.0 eV, J_{\text{H}}=J_{\text{pair}}\cong0.7 eV$. These
%interactions scales 
and can be 
obtained from constrained RPA calculations~\cite{Imada:2010}. 
%However, since we are here more interested in
%demonstrating how the optimization between the repulsive pairing
%interaction and the pair wave function allows for a unified picture
%of the FRG-results, we constrain ourselves to the above simplified
%interactions and vary them in a wide parameter regime.

Details of the electronic structure such as the FS topology and,
more specifically, the presence of the $(\pi, \pi)$-pocket are
crucial for the SC state
\cite{Mazin:2008,Kuroki:2008,Graser:2009,Stanev:2008,Chubokov:2008,Wang:2009,Platt:2009,Thomale:2009,Zhai:2009,Wang:2010,Chubokov:2009,Thomale:2010,Kemper:2010}.
Using FRG on the above 5-band (Fe $d$-orbital) model of the Fe-based
SC with orbital interactions as in Eq.~\ref{h0} and~\ref{hint}, we have recently
found that the gap on the e-pockets can undergo a nodal transition
if the h-pocket at $(\pi, \pi)$ is absent~\cite{Thomale:2010}.
Similar conclusions have been reached by Maiti and Chubukov using a
parquet RG analysis~\cite{Maiti:2010}. On the basis of
RPA calculations, Kuroki et al.~\cite{Kuroki:2008} using a similar
Hamiltonian as in Eqs.~\ref{h0} and~\ref{hint} have already argued that the
$(\pi, \pi)$-pocket is sensitive to the lattice structure
("pnictogen height") and crucial for the gap structure. Kemper et
al.~\cite{Kemper:2010}, within again a 5-orbital RPA theory, further
substantiated these conclusions. In particular, they demonstrated
the sensitive dependence of the SC state to aspects of the
electronic structure such as the FS topology and the FS orbital weights.

%In view of the sensitivity of the SC gap features to details of the
%electronic structure as well as to details of the competing
%many-body interactions in Eq. (2), we ask whether there is a kind of
%"ordering principle" helping us to further improve our understanding
%of this sensitivity. It is our belief that for studying these
%competitions, the FRG is an appropriate tool:

\begin{figure}[t]
  \begin{minipage}[l]{0.95\linewidth}
%    \psfrag{a}{$0$}
%    \psfrag{b}{$\uparrow$}
%    \psfrag{c}{$\downarrow$}
%    \psfrag{D}{\small$\Delta_3$}
    \includegraphics[width=\linewidth]{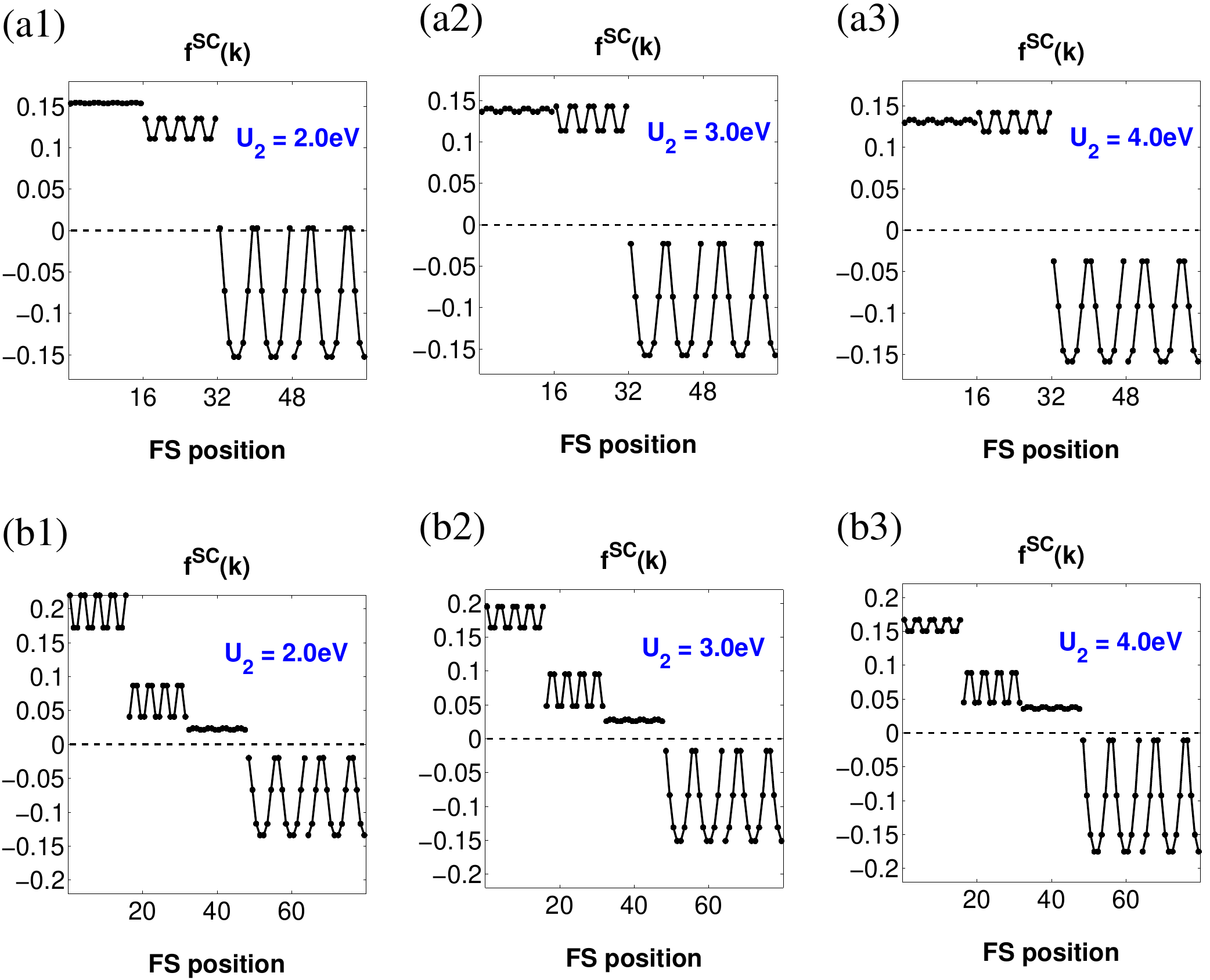}
  \end{minipage}
  \caption{(Color online). Scan of $U_{2}$. Upon increase, more and more relevant scattering contributions from the 3rd hole pocket (M) in the 5pFS scenario (row (b)), to parts of the electron pockets of different orbital content increase the anisotropy. In the 4pFS scenario (row (a)), the anisotropy decreases due to a weakening of the previously dominant scattering between the peaks and valleys of the e-pockets.}
\label{su-pic2}
\vspace{-0pt}
\end{figure}

\begin{figure}[t]
  \begin{minipage}[l]{0.95\linewidth}
%    \psfrag{a}{$0$}
%    \psfrag{b}{$\uparrow$}
%    \psfrag{c}{$\downarrow$}
%    \psfrag{D}{\small$\Delta_3$}
    \includegraphics[width=\linewidth]{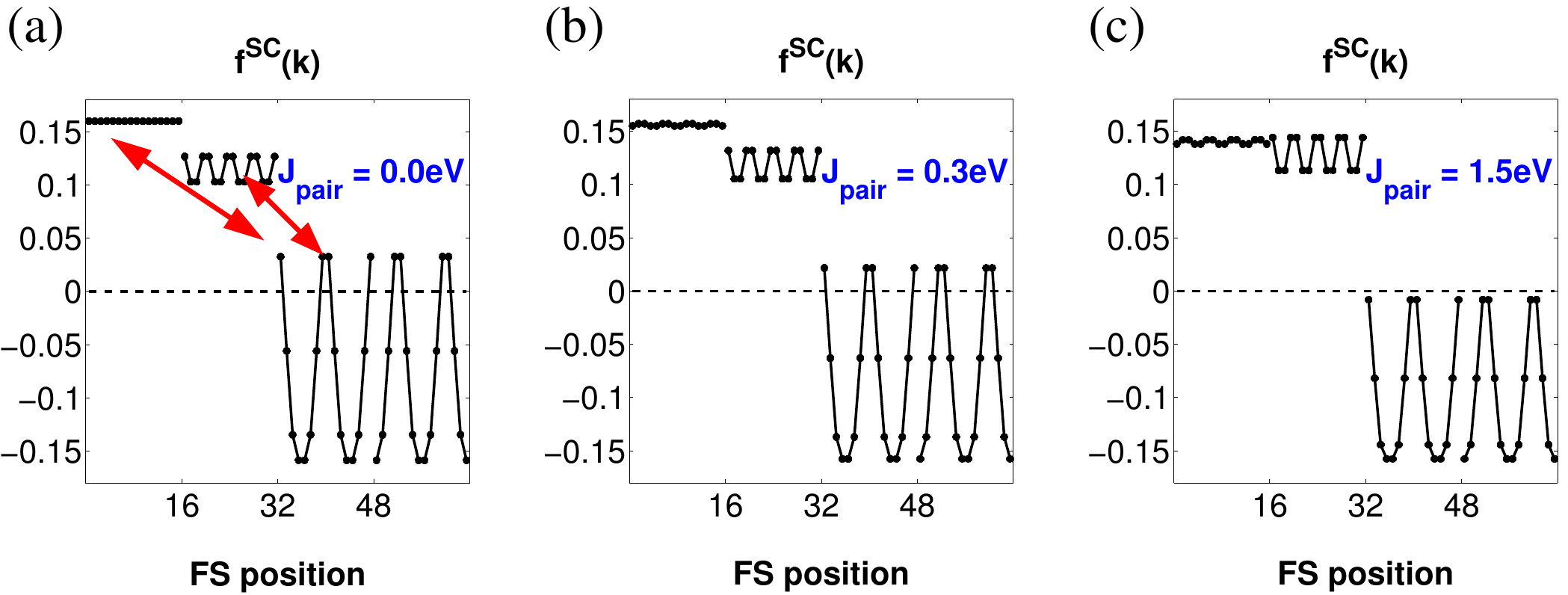}
  \end{minipage}
  \caption{(Color online). Scan of $J_{\text{pair}}$. The increase of
     pair-hopping helps to drive a gap for the 4pFS.}
\label{su-pic4}
\vspace{-0pt}
\end{figure}

In the FRG~\cite{Wang:2009,Platt:2009,Thomale:2009,Thomale:2010,Honerkamp:2001}, one starts from the "bare" many-body interaction (Eq.
(2)) in the Hamiltonian and the pairing is dynamically generated by
systematically integrating out the high-energy degrees of freedom
including important fluctuations (magnetic, SC, screening, vertex
corrections) on equal footing. This differs from the RPA which takes
right from the outset a magnetically driven SF-type of pairing
interaction. %In the RPA-calculation of Kuroki et al.
For a given instability characterized by some order
parameter $\hat{O}_{{\bf k}}$, the 4-point function (4PF) $V_{\Lambda}({\bf k}_1,{\bf
  k}_2,{\bf k}_3,{\bf k}_4)$ in the particular ordering
channel can be written in shorthand notation as $\sum_{{\bf k},{\bf p}}V_{\Lambda}({\bf
k},{\bf p}) [\hat{O}^{\dagger}_{{\bf k}}
\hat{O}^{\phantom{\dagger}}_{{\bf p}}]$~\cite{Zhai:2009}.
Accordingly, the 4PF $V_{\Lambda}({\bf k},{\bf -k},{\bf p},{\bf -p}))$ in the Cooper channel can be decomposed into
different eigenmode contributions~\cite{Wang:2009,Thomale:2010}
\begin{equation}
V^{\text{\text{SC}}}_{\Lambda} ({\bf k},{\bf p})= \sum_i c_i^{\text{\text{SC}}}(\Lambda) f^{\text{\text{SC}},i}({\bf k})^* f^{\text{\text{SC}},i}({\bf p}),
\label{decomp}
\end{equation}
where $i$ is a symmetry decomposition index,
% that agrees with the number of
%discretized momentum points (Fig.~\ref{pic2}b and~\ref{pic3}c),
and the leading instability of that channel corresponds to an
eigenvalue $c_1^{\text{\text{SC}}}(\Lambda)$ first diverging under the flow
of $\Lambda$. $f^{\text{\text{SC}},i}(\bf{k})$ is the SC form factor of
pairing mode $i$ which tells us about the SC pairing symmetry and
hence gap structure associated with it. In FRG, from the final
Cooper channel 4PFs, this quantity is computed along the discretized
Fermi surfaces (as shown in Fig. \ref{su-pic1}), and the
leading SC instabilities are plotted in
Figs.~\ref{su-pic1} to~\ref{su-pic4}. If not stated differently,
the interaction parameters not specified in the plots are kept fixed
at the representative setup $U_1 = 3.5 eV, U_2 = 2.0 eV,J_{\text{H}}=J_{\text{pair}}=0.7 eV$.

We first investigate the behavior upon the variation of the
intraorbital interaction scale $U_{1}$ (Fig.~\ref{su-pic1}). For
small values of $U_1 (U_1=U_2=2eV)$ in Fig. \ref{su-pic1}a, the
$s_{\pm}$-sign change is induced by the "optimization principle"
between the h-pockets around the $\Gamma$-point and the
e-pockets around the $X$
% $((\pi, 0)(0, \pi))$ 
points. By
increasing $U_{1}$, the 4pFS system  develops a pronounced gap
anisotropy at the electron pockets, which eventually leads to gap
nodes at $U_{1}\sim 3eV$ (Fig.~\ref{su-pic1}a). This behavior is
due to an enhanced $U_{1}$ repulsion within the $d_{X^2-Y^2}$
orbitals, which amplifies the pair scattering within the
$d_{X^2-Y^2}$-dominated parts of one electron pocket to the other
(see green arrow in Fig. \ref{su-pic1}a for $U_1=5eV$). The above mentioned
optimization requirement between a repulsive interaction and the
pair wave function then favors an increased gap anisotropy between the peak and
the valley of the SC gap on the e-pockets, which eventually
yields a sign change.

In the 5pFS scenario (Fig.~\ref{su-pic1}b), the additional
$M$-pocket, which exclusively carries $d_{X^2-Y^2}$ orbital weights,
also generates pair scatterings due to $U_{1}$ to the
$d_{X^2-Y^2}$-dominated parts of the electron pockets (see magenta
arrows in Fig.~\ref{su-pic1}b). This pushes down the
peaks (tips) of the SC-gap on the electron pockets, on the basis of
the same orthogonality argument (magenta arrow in Fig.~\ref{su-pic1}b).
%By increasing $U_{1}$, the As-based system {\color{red}develops a larger gap on the electron pockets}. This agrees with the %mechanism elaborated on above. The $(\pi,\pi)$ hole pocket contributes a scattering {\color{red}process} driven by $U_{1}$ that %increases the minimal gap on the upper tips of the electron pockets.
% At the same time, the increased $U_{\text{intra}}$ induces a larger gap driven by $U_{\text{intra}}$. However, as the increase of the front tip is compensated for by the likes increased front tip scattering between the two electron pockets, the total anisotropy of the SC form factor is slightly increased.
%For the P-based scenario with only 4 pockets, only the only $U_{\text{intra}}$-driven $d_{X^2-Y^2}$ scattering contributions are %given by the scattering between the front tips of the electron pockets, which favors the nodal propensity of the SC form factor. As %a consequence, the increase of $U_{\text{intra}}$ favors the nodal propensity.

As the inter-orbital interaction $U_{2}$ is increased
(Fig.~\ref{su-pic2}), the main change is that the significance of
the $d_{X^2-Y^2}$ scatterings driven by $U_{1}$ is slightly lowered.
As the interorbital scattering phase
space becomes important, the orbital distribution along
the pockets determines more and more the behavior. For the 5pFS
scenario (row (b) in Fig.~\ref{su-pic2}), this gives an
increased e-pocket anisotropy and a smaller e-pocket gap.
% where the minimal electron pockets gap is not coupled anymore to the size of the $(\pi,\pi)$ hole pockets as it is the case for the $U_{\text{intra}}$ dominated scenario.
For the 4pFS scenario, the nodal propensity is significantly
reduced, as the previously decisive scattering between the peaks and
the valleys of the electron pockets becomes less relevant.

%\begin{figure}[h]
%  \begin{minipage}[l]{0.95\linewidth}
%%    \psfrag{a}{$0$}
%%    \psfrag{b}{$\uparrow$}
%%    \psfrag{c}{$\downarrow$}
%%    \psfrag{D}{\small$\Delta_3$}
%    \includegraphics[width=\linewidth]{jh}
%  \end{minipage}
%  \caption{(Color online). Scan of $J_{\text{H}}$ for the P-situation which verifies that there is no significant change for the SC from factor in a large parameter regime.}
%\label{su-pic3}
%\vspace{-0pt}
%\end{figure}
As $J_{\text{H}}$ is increased, similar to
Wang et al.~\cite{Wang:2010}, we observe that the anisotropy on the
e-pockets is enhanced, again because of the general orthogonality
requirement between the repulsive interaction and the SC pair state,
which applies both to 5pFS and 4pFS scenarios. Within a reasonable
parameter range up to $\sim 1eV$, the modification of the SC form
factor is comparably small. %For
%rather large values of $J_{\text{H}}$ (not justified from DFT-type
%of calculations) again the general orthogonality principle is
%obvious: $J_{\text{H}}$ creates repulsive interactions in the RG
%flow which, in the P-based case (row (a)), try to further open up
%the difference between the tip and the valley of the gap function up
%to a nodal behavior.

%For even larger (and probably physically not relevant) bare values of $J_{\text{H}}$, differences between the P-based scenario and %the As-based scenario appear.
%One reason certainly is that all 4 pockets in the P-based scenario have changing orbital contributions along the pockets and thus %cooperate with the tendency of $J_{\text{H}}$ to favor anisotropy, while the scattering contribution from the additional 5th pocket %in the As-based scenario counteract this tendency.
On the bare level, $J_{\text{pair}}$ gives a positive semidefinite
contribution to the Cooper channel. As a consequence, it increases
the absolute Cooper channel scale and leads to an increase of the
$\Lambda$-divergence scale of the SC instability (not shown here).
While the SC form factor for the 5pFS scenario remains essentially
unchanged, we observe some
decrease of nodal propensity in the 4pFS scenario (see
Fig.~\ref{su-pic4}). There, the $J_{\text{pair}}$ scattering
contribution between the $\Gamma$ hole pockets and the tips of the
electron pockets play a similar role as the scattering contribution
of the 5th hole pockets in the 5pFS scenario. However, this
scattering takes place now between
$d_{XZ}\rightarrow d_{X^2-Y^2}$ and $d_{YZ}\rightarrow d_{X^2-Y^2}$,
orbitals (red arrows in Fig.~\ref{su-pic4}).
%In summary, the
%wide-ranged parameter studies confirm the picture developed in our
%Letter, that a $U_{1}$-dominated interaction setup, in combination
%with a 5- and 4-pocket scenario, is responsible for the differences
%between the As-based and the P-based 1111 compounds.
%Otherwise, not only the existence or non-existence of the third hole pocket, but likewise other detailed ratios of interaction %parameters are essential to understand the resulting SC phase.

%SUMMARY

In summary, we have demonstrated  the usefulness of the optimization
principle to provide a more universal characterization of
gap anisotropies in multi-band SC. The optimization minimizes the
Coulomb repulsion, which is needed because of frustration (such as
occurring in the $s_{\pm}$-channel in our example), when more than
two FS-sheets are contributing to the pairing interaction. Using FRG
for two generic FS topology setups of the pnictides, we show
that the optimization principle relates the gap anisotropies and
their nodeless versus nodal behavior in a rather transparent way to
the multi-orbital band structure, FS topology, and
interactions.

\section{Acknowledgements}
Useful discussions are gratefully acknowledged with 
D. J. Scalapino, C. Honerkamp and B. A. Bernevig. The work 
was supported by DFG-SPP 1458/1 (CP, RT and WH) and by the Bavarian KONWIHR Program. RT is
supported by the Humboldt Foundation.


\begin{thebibliography}{99}
\bibitem{Anderson:1987} P. W. Anderson, Science {\bf235}, 1196
(1987)

\bibitem{Scalapino:2007} See, for example, D. J. Scalapino, Chapt.
13 in "Handbook of High Temperature Superconductors" editors J. R.
Schrieffer and J. S. Brooks, Springer (2007), cond-mat/0610710 and
refs. therein.

\bibitem{Hackl:2010} For a recent review, see also "Towards a Better
Understanding of Superconductivity at High Transition Tempereatures"
editors R. Hackl and W. Hanke, Eur. Phys. J. Special Topics
{\bf188}, 3 (2010)

\bibitem{Mazin:2008} I. I. Mazin, D. J. Singh, M. D. Johannes, and M. H. Du, Phys. Rev. Lett. {\bf101}, 057003 (2008)

\bibitem{Kuroki:2008} K. Kuroki,  H. Usui,S. Onari, R. Arita, and H. Aoki, Phys. Rev. B {\bf79},
224511 (2009)

\bibitem{Graser:2009} S. Graser, T. A. Maier, P. J. Hirschfeld, and
D. J. Scalapino, New Journal of Physics {\bf11}, 025016 (2009) and
Phys. Rev. B {\bf79}, 224511 (2009)

\bibitem{Stanev:2008} V. Stanev, J. Kang, and Z. Tesanovic, Phys.
Rev. B {\bf78}, 184509 (2008)

\bibitem{Chubokov:2008} A. V. Chubukov, D. V. Efremov, and I.
Eremin, Phys. Rev. B {\bf78}, 134512 (2008)

\bibitem{Wang:2009} F. Wang, H. Zhai, Y. Ran, A. Vishwanath, and
D.-H. Lee, Phys. Rev. Lett {\bf102}, 1047005 (2009)

\bibitem{Platt:2009} C. Platt, C. Honerkamp, and W. Hanke, New J.
Phys. {\bf11}, 055058 (2009)

\bibitem{Thomale:2009} R. Thomale, C. Platt, J. Hu, C. Honerkamp,
and B. A. Bernevig, Phys. Rev. B {\bf80}, 180505 (2009)

\bibitem{Zhai:2009} H. Zhai, F. Wang, and D.-H. Lee, Phys. Rev. B
{\bf80}, 064517 (2009)

\bibitem{Wang:2010} F. Wang, H. Zhai, and D.-H. Lee, Phys. Rev. B
{\bf81}, 184512 (2010)

\bibitem{Chubokov:2009} A. V. Chubukov, M. G. Vavilov, and A. B.
Vorontsov, Phys. Rev. B {\bf80}, 140515 (2009)

\bibitem{Thomale:2010} R. Thomale, C. Platt, W. Hanke, and B. A.
Bernevig, arXiv:1001.3599 (2010)

\bibitem{Kemper:2010} A. F. Kemper, T. A. Maier, S. Graser, H.-P.
Cheng, P. J. Hirschfeld, and D. J. Scalapino, New Journal of Physics {\bf12}, 073030 (2010)

\bibitem{17} 
%At first glance, this
%similarity of the gap structure using a quite different technique,
%such as the RPA versus the RG may appear surprising.  
This similarity is partly based on the fact that the anisotropic wave function completely
avoids the strongly repulsive (short-range) part of the Coulomb
interaction (i.e. $U \sum_k\langle c_{k\dagger}c_{-k\dagger}\rangle
=0$), in a single-band Hubbard-U model, or, to a large extent for a
2-pocket toy model~\cite{Mazin:2009}. This is crucial for
RPA-treatments, where the bare repulsion U appears also in the
pairing channel. In the RG treatments for a 2-pocket situation the
corresponding intra-pocket repulsive part of the Coulomb interaction
is renormalized in the flow eventually even changing its sign to
become attractive. Yet, the pairing, i. e. the corresponding flow,
is dominated by the repulsive inter-pocket pair scattering with wave
vector $(\pi, 0)$ connecting the e- and h-pocket
\cite{Chubokov:2008,Platt:2009}. Thus, in the RG, the pair-wave
function is "optimized", i.e. gives the largest pairing eigenvalue
(see below), again for an $s_{\pm}$-form.
%Anderson et al., Science, {\bf317}, 1705(2007)

\bibitem{18} Some of the arguments in our Letter for the multi-pocket scenarios are
independent of the method chosen for extracting the pairing and
similar to arguments found e.g. in Ref.~\cite{Maiti:2010,Maier:2010}. However,
the "optimization", i.e. balancing the attractive $s_{\pm}$
interaction and the Coulomb repulsion (frustration) is different in
RG versus RPA treatments (see also Ref.~\cite{17} and \cite{Maiti:2010}).
%D. J. Scalapino, Phys. Reports {\bf250},329-365 (1995)

\bibitem{Kim:2010} For a recent experimental work and refs. therein
see e.g. H. Kim et al., arXiv:1008.3251v3 (2010)

\bibitem{Imada:2010} For a recent review see M. Imada and T. Miyake,
ch.5.2.6,  J. Phys. Soc. Jpn. {\bf79}, 112001 (2010)

\bibitem{Maiti:2010} S. Maiti and A. B. Chubukov, arXiv:1010.0984v2
(2010)

\bibitem{Maier:2010} T. A. Maier, S. Graser, D. J. Scalapino, and P.
Hirschfeld, Phys Rev. B {\bf79}, 224510 (2009)

\bibitem{Honerkamp:2001} C. Honerkamp, M. Salmhofer, N. Furukawa,
and T. M. Rice, Phys. Rev. B {\bf63}, 035109 (2001)

%\bibitem{Zhang:2009} Of course, in our parameter sweeps, we assume
%following J. Zhang et al., Phys. Rev. B {\bf79}, 220502 (R) (2009),
%that we have not a spin-rotationally invariant situation.

\bibitem{Mazin:2009} I. Mazin and J. Schmalian Physica C, {\bf469}, 614
(2009)
\end{thebibliography}
\end{document}